\documentclass[smallextended]{svjour3}
\usepackage{booktabs} 
\usepackage{array}
\pdfoutput=1
\usepackage{romannum}
\usepackage{balance}
\usepackage{multirow}
\usepackage{graphicx}
\usepackage{url}
\newcommand{\conclusion}[1]{%
	\begin{center}\noindent\thicklines\setlength{\fboxsep}{8pt}\fbox{\begin{minipage}{4.4in}\textit{\textbf{#1}}\end{minipage}}\end{center}} 
\usepackage{listings,xcolor}
\begin{document}
\title{MSRBot: Using Bots to Answer Questions from Software Repositories}

\author{Ahmad Abdellatif \and Khaled Badran \and Emad Shihab}

\institute{Ahmad Abdellatif, Khaled Badran and Emad Shihab\at
	Data-Driven Analysis of Software (DAS) Lab\\
	Department of Computer Science and Software Engineering\\
	Concordia University\\
	Montreal, Quebec, Canada\\
	  \email{\{a\_bdella, k\_badran, eshihab\}@encs.concordia.ca}
}

\maketitle
\begin{abstract}
Software repositories contain a plethora of useful information that can be used to enhance software projects. Prior work has leveraged repository data to improve many aspects of the software development process, such as, help extract requirement decisions, identify potentially defective code and improve maintenance and evolution. However, in many cases, project stakeholders are not able to fully benefit from their software repositories due to the fact that they need special expertise to mine their repositories. Also, extracting and linking data from different types of repositories (e.g., source code control and bug repositories) requires dedicated effort and time, even if the stakeholder has the expertise to perform such a task.

Therefore, in this paper, we use bots to automate and ease the process of extracting useful information from software repositories. Particularly, we lay out an approach of how bots, layered on top of software repositories, can be used to answer some of the most common software development/maintenance questions facing developers. We perform a preliminary study with 12 participants to validate the effectiveness of the bot. Our findings indicate that using bots achieves very promising results compared to not using the bot (baseline). Most of the participants (90.0\%) find the bot to be either useful or very useful. Also, they completed 90.8\% of the tasks correctly using the bot with a median time of 40 seconds per task. On the other hand, without the bot, the participants completed 25.2\% of the tasks with a median time of 240 seconds per task. Our work has the potential to transform the MSR field by significantly lowering the barrier to entry, making the extraction of useful information from software repositories as easy as chatting with a bot.

\end{abstract}


\section{Introduction}
\label{sec:intro}

Software repositories contain an enormous amount of software development data. This repository data is very beneficial, and has been mined to help extracts requirements (e.g.,~\cite{Ali2013TSE,Mordinyi2017REW}), guides process improvements (e.g.,~\cite{Gupta_MSR2014,Siddiqui_BDA2018}) and improves quality (e.g.,~\cite{Hassan2008FOSM,Khomh2015ESE}). However, we argue that even with all of its success, the full potential of software repositories remains largely untapped. For example, recent studies presented some of the most frequent and urgent questions (e.g., ``Where do developers make the most mistakes?'' and ``Which mistakes are the most common?'') that software teams struggle to answer~\cite{Begel:2014:ATQ:2568225.2568233}. Many of the answers to such questions can be easily mined from repository data.

Although software repositories contain a plethora of data, extracting useful information from these repositories remains to be a tedious and difficult task~\cite{Banerjee_BIGDSE2015,bankier2014}. Software practitioners (including developers, project managers, QA analysts, etc.) and companies need to invest significant time and resources, both in terms of personnel and infrastructure, to make use of their repository data. Even getting answers to simple questions may require significant effort.

More recently, bots were proposed as means to help automate redundant development tasks and lower the barrier to entry for information extraction~\cite{Storey:2016:DDP:2950290.2983989}. Hence, recent work laid out a vision for how bots can be used to help in testing, coding, documenting, and releasing software~\cite{Beschastnikh_ASE2017,Xu:2017:AAG:3155562.3155650,Tian_ASE2017,Wessel_2018CSCW}. To bridge the gap between the visionary works and practicality, and to bring those visionary works to life, we devise a framework of using bots over software repositories. Although different bots have been developed for the software engineering domain, no prior work has applied bots to answer the developers questions using the stored data from software repositories.

Although it might seem like applying bots on software repositories is the same as using them to answer questions based on Stack Overflow posts, the reality is there is a big difference between the two. One fundamental difference is the fact that bots that are trained on Stack Overflow data can provide general answers, and will never be able to answer project-specific questions such as ``how many bugs were opened against my project today?''. Also, we would like to better understand how bots can be applied on software repository data and highlight what is and what is not achievable using bots on top of software repositories.

Therefore, our goal is to design and build a bot framework for software repositories and perform a case study to examine its efficiency and highlight the challenges facing our framework. The approach contains five main components, a \textbf{user interaction} component, meant to interact with the user; \textbf{entity recognizer} and \textbf{intent extractor} components, meant to process and analyze the user's natural language input; a \textbf{knowledge base} component, that contains all of the data and information to be queried; and a \textbf{response generator} component, meant to generate a reply message that contains the query's answer and return it to the user interaction component. To evaluate our bot approach, we add support for 15 of the most commonly asked questions by software practitioners mentioned in prior work~\cite{4497212,6062080,Sharma:2017:DWA:3100317.3100333,Begel:2014:ATQ:2568225.2568233,Fritz:2010:UIF:1806799.1806828}. To evaluate our framework, we perform a case study with 12 participants using the Hibernate and Kafka projects. In particular, we asked those participants to perform a set of tasks using the bot then evaluate it based on its replies. We examine the bot in terms of its effectiveness, efficiency, and accuracy and compare it to a baseline where the survey participants are asked to do the same tasks without using the bot. We also perform a post-survey interview with a subset of the survey participants to better understand the strengths and areas of improvements of the bot approach. 

Our results indicate that bots are useful (as indicated by 90.0\% of answers), efficient (as indicated by 84.17\% of answers) and accurate (as indicated by 90.8\% of tasks) in providing answers to some of the most common questions. In comparison to the baseline, the bots significantly outperform the manual process of finding answers for their questions (the survey participants were able to only answer 25.2\% of the questions correctly and took much longer to find their answers). Based on our post-survey interviews with the participants, we find that bots can be improved if they enable users to perform deep-dive analysis and help compensate for user errors, e.g., typos. Based on our results, we believe that applying bots on software repositories has the potential to transform the MSR field by significantly lowering the barrier to entry, making the extraction of useful information from software repositories as easy as chatting with a bot.

In addition to our findings, the paper provides the following contributions:
\begin{itemize}
	\item[$\bullet$] To the best of our knowledge, this is the first study to use bots on software repositories. Also, our framework allows project stakeholders to extract repository information easily using natural language.
	\item[$\bullet$] We perform an empirical study to evaluate our bot framework and compare it to a baseline. Also, we provide insights on areas where bot technology/frameworks still face challenges in being applied to software repositories.
	\item[$\bullet$] We make our framework implementation \cite{MSRBot_Linkonline} and datasets \cite{BotData} publicly available in an effort to accelerate future research in the area.
\end{itemize}

\textbf{Paper Organization.} The rest of the paper is organized as follows. Section 2 provides background on the bot and discusses the related work to our study. We detail our framework and its components in Section 3. We explain the questions supported by the bot and questionnaire survey to evaluate our framework in Section 4. In section 5, we report our findings, detailing the usefulness, speed, and accuracy of the bot. Section 6 discusses the bot evaluation and the implications of our results. Section 7 discusses the threats to validity, and section 8 concludes the paper.

\section{Background}
\label{sec:relatedwork}

In this section, we provide a brief background of bots and their roles in the software development process. Also, we discuss the work that is most related to ours. We divide the prior work into two main areas; work related to the visions for the future use of bots, and answering developers' questions.

\noindent\textbf{Software Bots.}
Storey and Zagalsky defined bots as tools that perform repetitive predefined tasks to save developer's time and increase their productivity~\cite{Storey:2016:DDP:2950290.2983989}. They outlined five areas where they see bots as being helpful: code, test, DevOps, support, and documentation bots. In fact, there exist a number of bots, mostly enabled by the easy integration in Slack that fit into each of the aforementioned categories, for example, Jirafe \cite{Jirafe_link}, a code bot that allows developers to report bugs easily. Similarly, Dr. Code is a test bot that tracks technical debt in software projects and many others that notify developers whenever a special action happens, e.g., Pagerbot. One key characteristic of these bots is that they simply automate a task, and do not allow developers or users to extract information (i.e., ask questions, etc.) that they need answers too. In our work, we design and evaluate a bot framework that is able to intelligently answer questions based on the repository data of a specific project.

Recently, software bots are getting more attention from Software Engineering researchers \cite{Storey:2016:DDP:2950290.2983989}. For example, Wyrich and Bogner~\cite{Wyrich_BotSE2019} implemented a bot that performs code refactoring and creates a pull request with the new changes. Paikari \textit{et al.}~\cite{aw} developed Sayme, a bot to help developers address code conflicts through its user interface. Moreover, bots are used to detect bugs and generate fixes for them \cite{vanTonder_BotSE2019,Monperrus_RepairnatorReportMagazine2019}. For example, Urli \textit{et al.}~\cite{Urli_ICSE2018} developed the Repairnator which is a repair bot for Java programs. The Repairnator keeps monitoring the CI of the project and in the case of a test failure, the bot reproduces the bug and generates a fix. On the other hand, developers can use recommendation bots to find reviewers for a pull request \cite{Kumar_BotSE2019}, static analysis tools to prevent bugs in their repositories \cite{Brown_BotSE2019}, and experts on a source code artifact \cite{Cerezo_BotSE2019}.

\noindent\textbf{Visions for the Future Use of Bots.}
In addition to the visionary work by Storey and Zagalsky~\cite{Storey:2016:DDP:2950290.2983989}, which presented a cognitive support framework in the bots landscape, a number of other researchers proposed work that laid out the vision for the integration of bots in the software engineering domain. In many ways, this visionary work motivates our bot framework. Acharya \textit{et al.}~\cite{Acharya_ICSE2016} proposed the idea of code drones, a new paradigm where each software artifact represents an intelligent entity. The authors outline how these code drones interact with each other, updating and extending themselves to simplify the developer's life in the future. They see the use of bots as key to bringing their vision to life. Similarly, Matthies \textit{et al.}~\cite{Matthies_BotSE2019} envisioned a bot that analyzes and measures the software project's data to help development teams track their project progress. On the other hand, researchers envisioned bots that generate fixing patches and validate the refactoring and bug fixes \cite{vanTonder_BotSE2019}, and explain those fixes to the developers \cite{Monperrus_BotSE2019}.

Beschastnikh \textit{et al.}~\cite{Beschastnikh_ASE2017} presented their vision of an analysis bot platform, called Mediam. The idea of Mediam is that developers can upload their project to GitHub and allow multiple bots to run on them, which will generate reports that provide feedback and recommendations to developers. The key idea of the vision is that bots can be easily developed and deployed, allowing developers quick access to new methods developed by researchers. Robillard \textit{et al.}~\cite{Robillard2017ICSME} envisioned a future system (OD3) that produces documentation to answer user queries. The proposed documentation is generated from different artifacts i.e. source code, Q\&A forums, etc.

The gap between these visionary works and industrial use, motivates our work in order to bridge this gap by bringing this visionary work to life. And, to support software developers that have different levels of technical knowledge in their daily tasks. However, our work differs in that we build a bot framework that extracts data from software repositories to allow developers to answer some of the most common questions they have - and our focus is more on how to build and evaluate such a framework.

\noindent\textbf{Answering Developer Questions.}
The work most related to ours is the work that built various approaches to help developers answer questions they may have. 
For example, Gottipati \textit{et al.}~\cite{Gottipati2011ASE} proposed a semantic search engine framework that retrieves relevant answers to user's queries from software threads. Bradley \textit{et al.}~\cite{Bradley_ICSE2018} developed a conversational developer assistant using the Amazon Alexa platform to reduce the amount of manual work done by developers (e.g., create a new pull request). Hattori \textit{et al.}~\cite{Hattori2013ASE} proposed a Replay Eclipse plugin, which captures the fine-grained changes and views them in chronological order in the IDE. Replay helps developers answer questions during the development and maintenance tasks. Treude \textit{et al.}~\cite{Treude2015TSE} proposed a technique that extracts the development tasks from documentation artifacts to answer developers' search queries.

Perhaps the work closest to ours, is the work that applied bots in the software engineering domain. Murgia \textit{et al.}~\cite{Murgia_MHI2016}, built a Stack Overflow answering bot to better understand the human-bot interaction. They deployed a bot, that impersonated a human and answered simple questions on Stack Overflow. Although their bot performed well, it faced some adoption challenges after it was discovered that it was a bot. Similar to Murgia, Xu \textit{et al.}~\cite{Xu:2017:AAG:3155562.3155650} developed AnswerBot, a bot that can summarize answers (extracted from Stack Overflow) related to developers' questions in order to save the developer time. Tian \textit{et al.}~\cite{Tian_ASE2017} developed APIBot, a framework that is able to answer developers' questions on a specific API using the API's documentation. APIBot is built on the SiriusQA assistant, however the authors' main contribution is the "Domain Adaption" component that produces the questions patterns and their answers. The authors ran APIBot on the Java 8 documentation and performed a study with 92 questions that developers and students asked about Java 8. The results showed that APIBot achieved 0.706 Hit@5 score (i.e., provided an answer in the top 5 answers) based on the questions and answers it was tested with.

Although our work is similar to the work mentioned above and shares a common goal to answer developers' questions, our work differs from the prior work in several ways. First, our work differs in that we apply bots on software repositories, which brings different challenges (e.g., having to process the repos and deal with various numerical and natural text data) than those compared to bots trained on natural language from Stack Overflow. However, we believe that our work complements the work that supports developer questions from Stack Overflow. Second, our work is fundamentally different, since our goal is to help developers interact and get information about their project from internal resources (i.e., their repository data, enabling them to ask questions such as ``who touched file x?''), rather than external sources such as Stack Overflow or API documentation that do not provide detailed project-specific information.

\noindent We believe that our work complements the previous work since, instead of using external sources of information to support developers, we focus on using an internal source of information (repositories data). Moreover, our work contributes to the MSR community by laying out how bots can be used to support software practitioners, allowing them to easily extract useful information from their software repositories. And, our findings shed light on issues that need to be addressed by the research community.

\section{MSRBot Framework}
\label{sec:approach}

\begin{figure}[t!]
	\centering
	\includegraphics[width=0.93\textwidth]{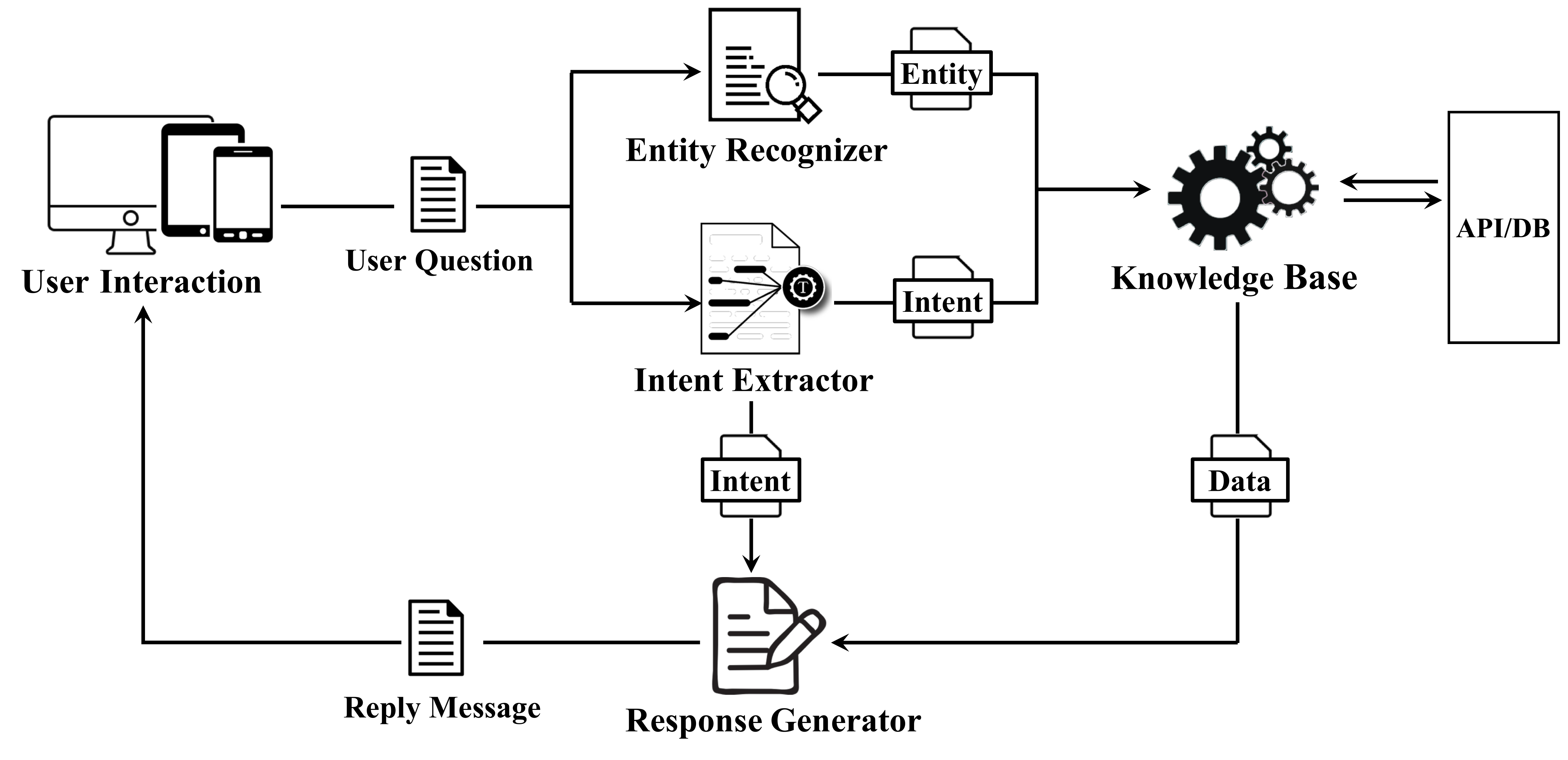}
	\caption{Overview of the MSRBot Framework and its Components' Interactions}
	\label{fig:Methodology Overview}
\end{figure} 

Our goal is to build a bot that users can interact with to ask questions (such as questions presented in Table \ref{table:ListOfQuestions}) to their software repositories. To enable this, our framework is divided into five main parts (as shown in Figure~\ref{fig:Methodology Overview}), namely 1) user interaction 2) entity recognizer 3) intent extractor 4) knowledge base and 5) response generator. In the following subsections, we detail each of these parts and showcase a working example of our framework.

\subsection{User Interaction}
Users of bot frameworks need to be able to effectively interact with their information. This can be done in different ways, e.g., through natural language text, through voice and/or visualizations. In addition to handling user input, the user interaction component also presents the output of the question to the user. This is done in the same window and appears as a reply to the user's question. Users are expected to pose their questions in their own words, which can be complicated to handle, especially since different people can pose the same question in many different ways. To help us handle this diversity in the natural language questions, we devise entity recognizer and intent extractor components, which extract structured information from unstructured language input (question) posted by the user. We detail those components in the next subsections.

\subsection{Entity Recognizer}
The entity recognizer component identifies and extracts a useful information (entity) that a user mentioned in the question using Named Entity Recognition (NER) \cite{Sang_2003ICNLL}. Also, it categorizes the extracted entities under certain types (e.g. city name, date, and time). There are two main NER categories: Rule-Based and Statistical NER. Prior work showed that statistical NER is more robust than the rule-based in extracting entities \cite{Ratinov_2009ICCNLL,Liu_2011MACL,mohit_2014Springer}. In the rule-based NER, the user should come up with different rules to extract the entities while in the statistical NER the user trains a machine learning model on an annotated data with the named entities and their types in order to allow the model to extract and classify the entities. The extracted entities help the knowledge base component in answering the user's question. For example, in the question: ``Who modified Utilities.java?'', the entity is ``Utilities.java'' which is of type ``File Name''.  Having the file name is necessary to know which file the user is asking about in order to answer the question correctly (i.e. bring the information of the specified file). However, knowing the file name (entity) is not enough to answer the user's question. Therefore, we also need an intent extractor component, which extracts the user's intention from the posed question. We detail this component in the next subsection.

\subsection{Intent Extractor}
The intent extractor component extracts the user's purpose/motivation (intent) of the question. In the last example, ``Who modified Utilities.java?'', the intent is to know the commits authors that modified the Utilities file. One of the approaches (e.g.,~\cite{Zamanirad_2017ASE}) to extract the intents is to use Word Embeddings, more precisely the Word2Vec model \cite{Mikolov_2013ICNIPS}. The model takes a text corpus as input and outputs a vector space where each word in the corpus is represented by a vector. In this approach, the developer needs to train the model with a set of sentences for each intent (training set). Where those sentences express the different ways that the user could ask about the same intent (same semantic). After that, each sentence in the training set is represented as a vector using the following equation:
\begin{equation}
\label{eq:wordvector}
\vec{Q} = \sum_{j=1}^{n}\vec{Q}_{wj}   \quad \textrm{Where} \  \vec{Q}_{wj} \  \epsilon \  VS
\end{equation}

where $\vec{Q}$ and $\vec{Q}_{wj}$ represent the word vector of a sentence and vector of each word in that sentence in the vector space $VS$, respectively. Afterwards, the cosine similarity metric~\cite{Jurafsky_2009SLP} is used to find the semantic similarity between the user's question vector (after representing it as a vector using Equation~\ref{eq:wordvector}) and each sentence's vector in the training set. The intent of the user's question will be the same as the intent of the sentence in the training set that has the highest score of similarity. The extracted intent is forwarded to the response generator component in order to generate a response based on the identified intent. Also, the intent is forwarded to knowledge base component in order to answer the question based on its intent. We explain this component in the next subsection. 

If the intent extractor is unable to identify the intent (low cosine similarity with the training set), it notifies the knowledge base and the response generator components, which respond with some default reply.

\subsection{Knowledge Base}
The knowledge base component is responsible for answering the user's questions (e.g. making an API call or querying a DB). It uses the passed intent from the previous component to map it with an API call or DB query that needs to be executed in order to get the answer of the question. And, it uses the extracted entities from the entity recognizer component as parameters for the query or call. For example, if a user asks the question ``Which commits fixed the bug ticket HHH-11965?'', then the intent is to get the fixing commits and the issue key ``HHH-11965'' is the entity. So, the knowledge base component uses the identified intent to retrieve the mapped query to the extracted intent and the entity as a parameter to that query. Therefore, the knowledge base component queries the database on the fixing commits (using SZZ \cite{Sliwerski_2005MSR}) that are linked to Jira ticket ``HHH-11965''. The component forwards the query's results to the response generator component to generate a reply message on the user's question. In case the intent extractor was unable to identify the intent, the knowledge base will do nothing and wait for a new intent and entities. Furthermore, the knowledge base component verifies the presence of the entities associated with the extracted intent and notifies the response generator in case of missing entities or unable to retrieve the data from the API. We describe the response generator component in the next subsection.

\subsection{Response Generator}
\label{subsec:response_generator}
The response generator component generates a reply message that contains the answer to the user's question and sends it to the user interaction component to be viewed by the user. The response is generated based on the question asked, and more specifically, the extracted intent of the question. Finally, it sends the generated message to the user interaction component.

In some cases, the bot may not be able to respond to a question due to lack of information (i.e., intents and entities). For example, if it is not possible to extract the intent, the response generator returns a default response:``Sorry, I did not understand your question, could you please ask a different question?''. And, in case of a missing entity, the response is ``Could you please specify the \textit{entity}?'' and it mentions the entity in the message (e.g., file name).

\begin{figure}[t!]
	\centering
	\includegraphics[width=1.0\textwidth]{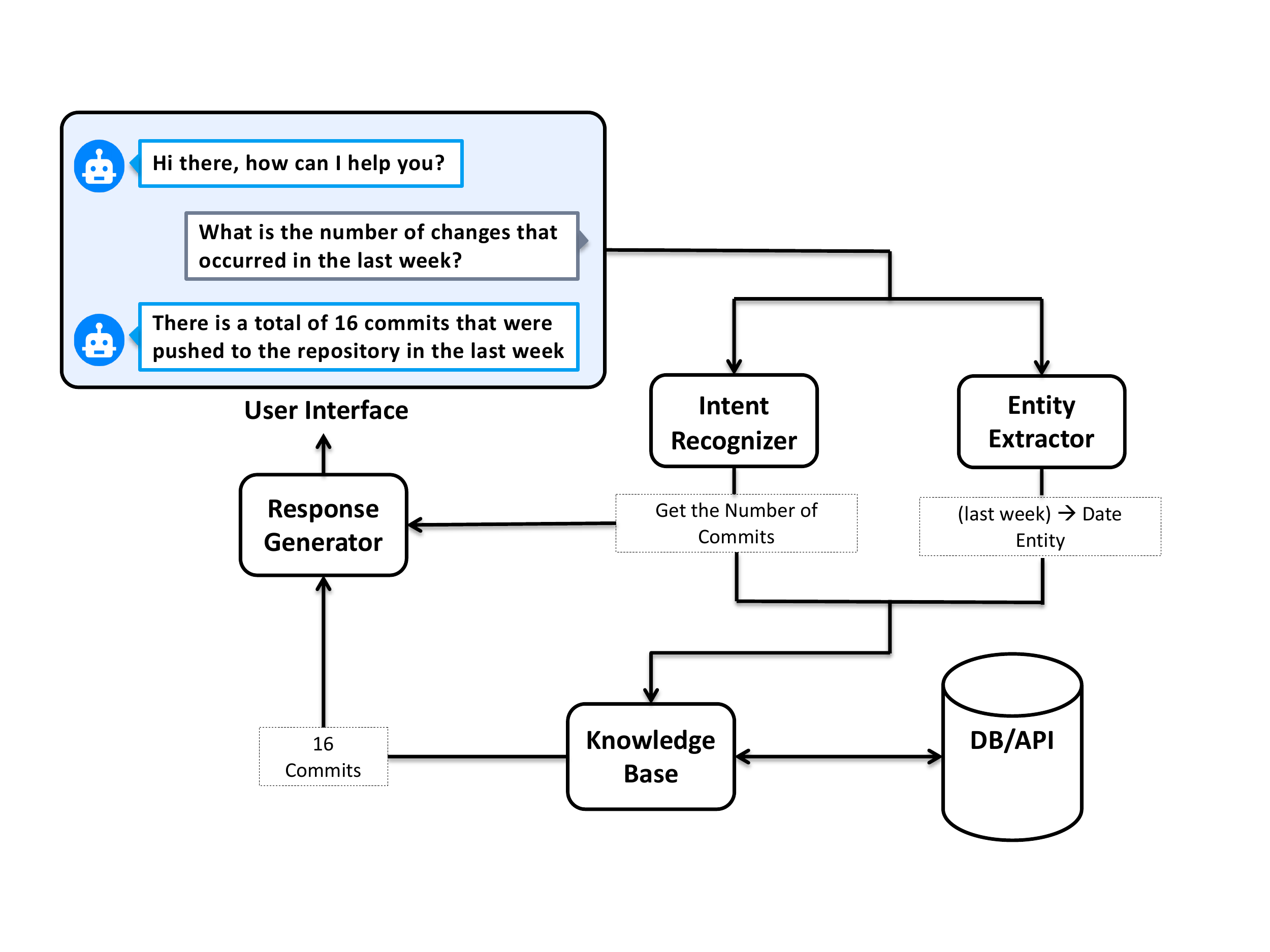}
	\caption{Example of MSRBot Framework Components Interactions to Answer User's Question}
	\label{fig:example}
\end{figure} 

It is worth mentioning that there are certain components that need to be customized to suit our approach. We customize the entity recognizer and intent extractor components to make our bot applicable on software repositories data. For example, we need to train the entity recognizer on the software engineering entities that are specific to software repositories (e.g., Jira tickets and commit hashes) to be able to identify those entities in the posed query. Also, our knowledge base component is specific to software repositories in a number of ways. First, it extracts and links the data from the source code and issue tracking repositories. Second, the knowledge base is customized to retrieve the answer to the user's queries based on the defined intents and entities. On the other hand, the remaining components are applicable to any type of software bots (e.g., user interface).

\subsection{Explanatory Example}

Putting everything together, we showcase an end-to-end example of the interactions between the components of our proposed framework to answer the user's question. In Figure 2, the user interacts with the bot through the user interface component to inquire about the number of changes that happened in the last week. Next, the user interface component forwards the question to the entity recognizer and intent extractor components to parse the user's query. The entity recognizer component identifies the entity `last week' (i.e., 21/01/2019 – 27/01/2019) of type `Date' in the query. On the other hand, the intent extractor component classifies the intent of the posed query as ``Get the Number of Commits''. Then, the results of the entity recognizer and intent extractor are sent to the knowledge base component to be processed. Also, the extracted intent is forwarded to the response generator component to generate the reply message. The knowledge base component uses the forwarded intent to retrieve the SQL query that is mapped to it while the entity is used as the parameter for the query. Next, the knowledge base executes the SQL query and retrieves the data from the database. Then, the knowledge base sends the query execution result (i.e., 16 commits) to the response generator. Finally, the response generator uses the results from the knowledge base and the forwarded intent from the intent extractor to generate the reply message (``There is a total of 16 commits that were pushed to the repository in the last week''), and sends it back to the user interface component to present the answer of the question to the user.

\section{Case Study Setup}
\label{sec:casesetup}

\begin{figure}[t!]
	\centering
	\frame{\includegraphics[width=0.87\textwidth]{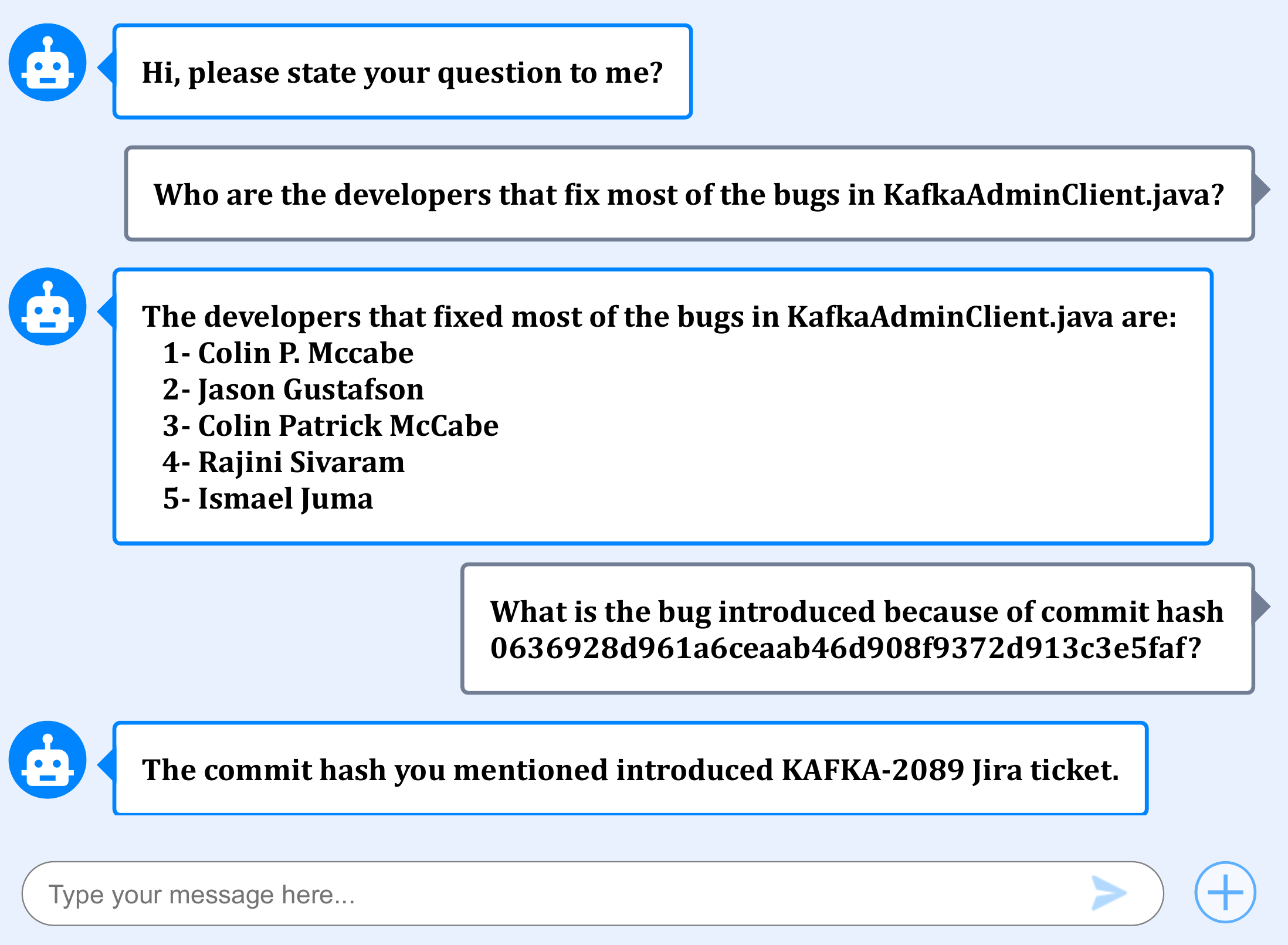}}
	\caption{An Example User Conversation with MSRBot}
	\label{fig:BotInterface}
\end{figure}

To determine whether using bots really helps answer tasks based on repository data, we perform an experiment with 12 participants using Hibernate-ORM and Kafka repositories. We built a web-based bot application that implemented our framework and had users directly interact with the bot through this web-application. A screenshot of our bot's interface is shown in Figure \ref{fig:BotInterface}. We divided the participants into two groups (each of 6 participants). Then, we sent emails that include links to the bot and online survey to both groups' members, asking each group to perform a set of tasks using the bot. Finally, we examine the bot in terms of its effectiveness, efficiency and accuracy and compare it to a baseline where both groups' members are asked to perform the same tasks without the bot. It is important to emphasize that each group performs the same set of tasks related to a specific repository. In other words, all members of group 1 performed the tasks using Hibernate-ORM, while the members of the group 2 used Kafka project.

To extract the intents and entities, we leveraged Google's Dialogflow engine~\cite{Dialogflow}. Dialogflow has a powerful natural language understanding (NLU) engine that extracts the intents and entities from a user's question based on a custom NLP model. Our choice to use Dialogflow was motivated by the fact that it can be integrated easily with 14 different platforms and supports more than 20 languages. Furthermore, it provides speech support with third-party integration and the provided service is free. These features make it easier to enhance our framework with more features in the future. 

Any NLU model needs to be trained. Therefore, to train the NLU, we followed the same approach as Toxtli et al.~\cite{Toxtli_2018CHI}. Typically, the more examples we train the NLU on, the more accurate the NLU model can extract the intents and entities from the users questions~\cite{DialogFlowTraining}. As a first step, we conducted a brainstorm session to create the initial training set which represents the different ways that the developers could ask for each intent in Table~\ref{table:ListOfQuestions}. Moreover, our bot can handle basic questions such as greeting users and asking general questions about the bot such as: ``How are you?''. Then, we used the initial set to train the NLU and asked two developers (each has more than 7 years of industrial experience) to test the bot for one week. During this testing period, we used the questions that the developers posed to the bot to further improve the training of the NLU. The training data is publicly available on~\cite{BotData}.

Although we use Dialogflow in our implementation, it is important to note that there exist other tools/engines that one can use such as Gensim~\cite{gensim_link}, Stanford CoreNLP~\cite{Manning_2014ACL}, Microsoft's LUIS~\cite{LUIS}, IBM Watson~\cite{IBMWatson}, or Amazon Lex~\cite{AmazonLex_link}.

To ensure that the usage scenario of the Bot is as realistic as possible, we asked the participants to perform a set of tasks using the bot. We used the questions that have been identified in the literature as being of importance to developers~\cite{Begel:2014:ATQ:2568225.2568233,Appendix12:online,Fritz:2010:UIF:1806799.1806828,Sharma:2017:DWA:3100317.3100333,6062080,4497212} to formulate the tasks in our experiment. The participants use the bot by posing any number of natural language questions needed to complete the task. Next, the participants evaluate the bot by answering a set of questions related to the task and bot's reply. To compare, we also ask the participants to perform the same tasks without the bot, which we call the baseline comparison. We detail all the steps of our case study in this section.

\begin{table*}[tbh]
\caption{List of Questions Examined by MSRBot and the Rationale for Their Inclusion}
\centering
	\begin{tabular} {@{}p{1.6in}p{2.2in}l@{}} 
		\toprule
		 \textbf{Question} & \textbf{Rationale} & \textbf{Ref.} \\ 
		\midrule
		
		Q1. Which commits fixed the \textbf{bug id}? & To refer the commit to a developer who is working on similar bug. &\cite{Begel:2014:ATQ:2568225.2568233}\\
		
		Q2. Which developer(s) fixes the most bugs related to \textbf{File Name}? & To know which developer(s) have experience in fixing bugs related to a specific component or file. &\cite{Begel:2014:ATQ:2568225.2568233}\\
		
		Q3. Which are the most bug introducing files? & To refactor the buggy files in the repository. &\cite{Begel:2014:ATQ:2568225.2568233}\\
		
		Q4. Who modified \textbf{File Name}? &To know which developer(s) worked on the file in order to ask them about a piece of code.& \cite{Sharma:2017:DWA:3100317.3100333,Fritz:2010:UIF:1806799.1806828}\\
		
		Q5. Which are the bugs introduced by commit \textbf{Commit Hash}? & To study the type of bugs introduced because of certain commit. &\cite{Begel:2014:ATQ:2568225.2568233}\\
		
		Q6. What is the number of commits in/on \textbf{Date}? & To track the progress of the team members at particular time (e.g., in the last day, week, month). & \cite{Fritz:2010:UIF:1806799.1806828}\\
		
		Q7. What commits were submitted on \textbf{Date}? & To know exactly what commits were done, to flag for review, testing, integration, etc. &\cite{Fritz:2010:UIF:1806799.1806828}\\
		
		Q8. What is/are the latest commit(s) to \textbf{File Name}? & Developers may want to know what are the last things that changed in a file/class to be up to date before they make modifications. &\cite{Sharma:2017:DWA:3100317.3100333,Fritz:2010:UIF:1806799.1806828,Appendix12:online}\\
		
		Q9. What are the commits related to \textbf{File Name}? & To track changes which are happening on the file which the developer is currently working on. Or to know the changes happening to a file that was abandoned by a developer. &\cite{6062080} \\
		
		Q10. What is/are the most common bug(s)? & The team wants the most important/common bug (registered as watchers) so it can be addressed. &\cite{Begel:2014:ATQ:2568225.2568233}\\
		
		Q11. What are the buggy/fixing commits that happened in/on \textbf{Date}? & To know the buggy or fixing commits happened at a particular time e.g. before release date. &\cite{Begel:2014:ATQ:2568225.2568233}\\
		
		Q12. How many bugs have the status/priority? & Quickly view the number of bug reports with a specific status (e.g., open) or priority (e.g., blocker) plan a fix for it. &\cite{Begel:2014:ATQ:2568225.2568233}\\
		
		Q13. Who is the author of \textbf{File Name}? & Developers who have questions about a specific file or class may want to speak to the person who created it. & \cite{4497212,Sharma:2017:DWA:3100317.3100333}\\
		
		Q14. Which developer(s) have the most unfixed bugs? & To determine the overloaded developers and possibly re-assign bugs to others on the team. &\cite{Begel:2014:ATQ:2568225.2568233}\\
		
		Q15. What is the percentage of bug fixing commits that introduced bugs in/on \textbf{Date}? & To study the characteristics of the fixing commits which happened at a certain time and induced bugs. &\cite{Begel:2014:ATQ:2568225.2568233}\\
		
		\bottomrule
	\end{tabular}
	\label{table:ListOfQuestions}
\end{table*}

\subsection{Questions Supported by the Bot}
\label{sec:questions}

To perform our study, we needed to determine a list of questions that our bot should support. To do so, we surveyed work that investigated the most commonly asked questions of software practitioners (mostly developers and managers). We found a number of notable and highly cited studies, such as the study by Begel and Zimmermann \cite{Begel:2014:ATQ:2568225.2568233,Appendix12:online}, which reported questions commonly asked by practitioners at Microsoft, the study by Fritz and Murphy~\cite{Fritz:2010:UIF:1806799.1806828} that conducted interviews with software developers to determine questions that developers ask and the study by Sharma \textit{et. al}~\cite{Sharma:2017:DWA:3100317.3100333}, which prioritized the importance of questions that developers ask. We also surveyed a number of other studies whose goal was not directly related to questions that developers ask, but which we found to be relevant for us to understand which questions we should ask (e.g.,~\cite{6062080,4497212}). In most of these studies, \emph{the authors reported that software practitioners are lacking support in answering these questions, which is exactly what our bot can help provide}. After going through the literature, we selected arbitrarily 15 questions that our bot can support and \emph{can be answered using repository data.} 

Table~\ref{table:ListOfQuestions} presents the questions we use in the case study, the rationale for supporting the question and the study where the question was mentioned/motivated from. Each question represents an intent and the bold words represent the entities in the question. For example, the user could ask Q11 as: ``What are the buggy commits that happened last week?'', then the intent is ``Determine Buggy Commits'' and the entity is ``last week''. It is important to emphasize that the bot's users can ask the questions in different ways other than what is mentioned in Table~\ref{table:ListOfQuestions}. In the last example the user can ask the bot ``What are the changes that introduced bugs on Dec 27, 2018'' where the intent remains the same although the question is asked in a different way and the entity is changed to a specific date (Dec 27, 2018).

Although we support 15 questions in our prototype at this time, it is important to note that the bot framework can support many more questions and we are extending it to do so now. We opted to focus on these 15 questions since our goal is to evaluate the bot in this research context and wanted to keep the evaluation manageable.

\subsection{Study Participants}

\begin{table}[t]
	\centering
	\caption{Participants' Knowledge on Version Control Repositories and Issue Tracking System} 
	\begin{tabular}{l|c|c}
		\toprule
		\multirow{2}{*}{\textbf{\begin{tabular}[c]{@{}c@{}} Likert \\  Scale\end{tabular}}} & \multicolumn{2}{c}{\textbf{Number of Participants}} \\ \cline{2-3} 
		& \textbf{\begin{tabular}[c]{@{}c@{}}Version Control \\Repositories (Git)\end{tabular}} & \textbf{\begin{tabular}[c]{@{}c@{}}Issue Tracking  \\System (Jira)\end{tabular}} \\  \midrule
		1 (No Knowledge) & 0 & 1 \\ 
		2 (Entry Level) & 2 & 4 \\ 
		3 (Intermediate) & 3 & 4 \\ 
		4 (Competent) & 5 & 1 \\ 
		5 (Expert) & 2 & 1 \\ \bottomrule
	\end{tabular}
	\label{Participants_Knowledge_Git_JIRA}
\end{table}

Once we decided on the 15 questions that we are able to support, we want to evaluate how useful the bot is. Since bots are meant to be used by real software practitioners, we decided to evaluate our bot through a user study.

Our user study involved 12 participants. For each participant, we asked them about their main occupation, background, software development experience and their knowledge of software repositories. All participants were graduate students (4 Ph.D. and 8 master students). Of the 12 participants, 75\% have more than 3 years of professional software development experience and 25\% have between 1-3 years of development experience. The participants' experience using Version Control Repositories (e.g., Git) and Issue Tracking System (e.g., Jira) are shown in Table \ref{Participants_Knowledge_Git_JIRA}.

We deliberately reached out to graduate students to conduct our user study for a few specific reasons. First, we knew that most of these graduate students (80\%) have worked in a professional software development environment in the past. Second, since this is one of the first studies using bots, we wanted to interview some of the participants in person, which provides us with invaluable feedback about aspects of using bots that may not come out in the user study. Expecting developers from industry, who are already busy and overloaded, dedicate this much time to our study would be difficult and if they did, we would not be able to go as deep in our study with them. Also, as prior work has shown, students can be a good proxy for what developers do in professional environments, especially if the participants are experienced and the technology under study is new, which is our case~\cite{Salman2015ICSE,Host2000ESE}. Lastly, the main goal of our case study is to evaluate the proposed framework, i.e., this is a judgment study rather than a sample study,  using students is completely valid. The real important factors for us is not the characteristics of the participants (students), rather the evaluation of the framework. Once we recruited our participants, we devised a questionnaire survey to evaluate the bot and baseline approaches. We detail our survey next.

\subsection{Questionnaire Survey}
We devised a survey that participants answer to help us understand the usefulness of the bot. To make the situation realistic, we mined the data from the Git and Jira repositories of the Hibernate-ORM and Kafka projects. Hibernate is a Java library that provides Object/Relational Mapping (ORM) support. And, Kafka is a Java platform that supports streaming applications. We setup our bot framework to be able to answer all the supported questions on Hibernate's and Kafka's repository data. There was no specific reason for choosing those projects as our case study, however, they did meet some of the most common criteria - they are large open source projects (Hibernate with 177 releases and Kafka with 97 releases) that uses Git and Jira, they have rich history (each has more than 6,500 commits, 8,800 bug reports), are popular amongst developers (each has more than 340 unique contributors) and have been studied by prior MSR-type studies (e.g.,~\cite{Ahmed2016MSR,Sawant2017ESE,Kabinna_2016MSR,Georgios_2017ECSA}).

Our survey was divided into three parts. The first section gathered information about the participants and asked questions related to experience, current role, and knowledge in mining software repositories, which is the information we presented in the section above. The second part of the survey was composed of 10 tasks that the participants are asked to perform. The task statements we gave the users all the needed information to complete the task. For example, the given task statement would say ``ask about the commits that fix HHH-6574'', and a user might use the bot to perform the task by asking: ``which commits fixed the bug id?''	In this case, the user is free to ask the question in any way they prefer, e.g., what are the fixing commits of HHH-6574 ticket? We provide a Jira ticket number that exists in Hibernate and Kafka, since we do not expect the user/participant to know this, however, someone related to any project is asking this question, s/he would indeed have such information. The remainder of the second part contained questions for the participants to evaluate the bot based on the answer they receive. We discuss the questions used to evaluate the bot in the next section.

In addition to asking the participants to complete the tasks using the bot, we also got them to perform the same tasks manually (without using the bot) to have a baseline comparison. For the baseline evaluation, we gave the exact tasks formulated from the questions shown in Table~\ref{table:ListOfQuestions} to the participants, so they know exactly what to answer to. The participants were free to use any technique they prefer such as writing a script, performing web searches, using tools (e.g., gitkraken~\cite{GitClienTool_link} and Jira Client~\cite{JiraClientTool_link}), executing Git/Jira commands, or searching manually for the answer in order to complete the tasks. Our goal was to resemble as close to a realistic situation as possible.

\subsection{Evaluating the Bot}
Bots are typically evaluated using factors that are related to both, accuracy and usability~\cite{Vasconcelos2017HFCS}. Particularly, this work suggested two main criteria when evaluating bots:
\begin{itemize}
	
	\item \textbf{Usefulness:} which states that the answer (provided by the bot) should include all the information that answers the question clearly and concisely \cite{Zamora2017HAI,Sankar2008SAICSIT}.

	\item \textbf{Speed:} which states that the answer should be returned in a time that is faster than the traditional way that a developer retrieves information~\cite{Zamora2017HAI}.

\end{itemize}

In essence, bot should help developers in performing their tasks and do this in a way that is faster than if you were not using the bot.
In addition to the two above evaluation criteria, we added another criteria, related to the accuracy of the answers that the bot provides. In our case, we define \textbf{accuracy} as the number of correctly completed tasks performed by the bot, where the task is marked as correct if the returned answer by the bot matches the actual answer the actual answer to the question \cite{Vasconcelos2017HFCS}. We formalize our case study with three research questions that are related to the three evaluation measures used, in particular we ask:

\begin{itemize}
	\addtolength{\itemindent}{1.1cm}
	\item[\textbf{RQ1:}] How useful are the bot's answers to users' questions?
	\item[\textbf{RQ2:}] How quickly can users complete their tasks using the bot?
	\item[\textbf{RQ3:}] How accurate are the bot's answers?	
\end{itemize}

To address \textbf{RQ1}, we ask two sub-questions. First, we ask whether the bot was able to return an answer in the first place (in some cases it cannot) and we ask the participants whether they consider that the answer returned is useful in answering the question posed on a five point Likert's scale (from very useless to very useful). 

For \textbf{RQ2}, we recorded the actual time the participants need to perform each of the tasks while using the bot through an online survey tool. For each task, we measure the time starting from the moment the participant is given the task until s/he submits the task. We use this time to quantitatively compare the time savings to the baseline, i.e., accomplishing the same tasks without using the bot. In addition, we ask the participants to indicate how fast the bot replies to their questions on a five point Likert's scale from very slow to very fast (to measure perceived speed). For the case of the baseline, we asked the participants to measure and report the time it took them to complete each task. We limited the maximum time to finish each task to 30 minutes when using the bot and in the baseline, i.e., in case they did not manage to get an answer within 30 minutes, we considered the task to be incomplete.

To address \textbf{RQ3}, we recorded all the interactions performed by the participants and, more importantly, the output of each bot's components including its responses. Then, we analyzed the tasks' results manually to determine if a task is completed correctly or not (by cloning the repositories and writing the scripts to answer the questions). For example, if the participant asks for the number of commits on a particular day, we would check if the returned answer was actually correct or not. This enables us to ensure that the entity recognizer, intent extractor, mapping process in the knowledge base, and the response generator components are working correctly. In the case of the baseline questionnaire, we asked the participants to provide us the answer of each task in the survey. This allowed us to determine the accuracy of the manually determined tasks. To ensure that the participants actually searched the answer for the asked task, we required that the participants briefly explain how they performed the task. Having this information also provided us with insight into how much work and what tools/techniques/commands practitioners typically use to perform such tasks. Finally, at the end of the survey, we added an optional field to allow the participants to write their comments or suggestions, if they have any.

To avoid overburdening the participants, we divided them into two groups, where each group has 6 participants and is given 10 tasks to perform on a certain repository. The tasks were formulated from 10 randomly selected questions from the list of questions supported by the bot. The questions that were selected to formulate the tasks are Q1, Q2, Q3, Q5, Q6, Q7, Q9, Q11, Q14, and Q15 in Table ~\ref{table:ListOfQuestions}. The first group performed the tasks on the Hibernate project while the members of the second group are instructed to do the same tasks on the Kafka project. Both groups (the Hibernate and Kafa groups) were asked to perform the tasks twice, once using the bot and another time without the bot (which we call the baseline). None of the participants knew the questions that the bot was trained on. This is to ensure they will use their own words when they are interacting with the bot and to monitor the questions that the participants ask the bot about. It is important to emphasize that each group received the same tasks in both, the baseline questionnaire and the bot-related questionnaire.

\section{Case Study Results}
\label{sec:results}

In total, the 12 participants asked the bot 165 questions (some developers asked more than 10 questions) to perform the assigned tasks \cite{BotData}. Of the 165 questions, we excluded 9 questions from our analysis because they were out of scope (e.g., ``What's your name?'', ``What language are you written in?''), as the main focus of this work is to study bots on software repositories. Therefore, all of the presented results are based on the remaining 156 questions that are relevant.

\subsection{RQ1: How useful are the bot's answers to users' questions?}

As mentioned earlier, one of the first criteria for an effective bot is to provide its users with useful answers to their questions. Evaluating a bot by asking how useful its answers were commonly used in most bot-related research (e.g.~\cite{Xu:2017:AAG:3155562.3155650,Zamora2017HAI,Feng2006ICIUI}).

\begin{figure}[t!]
	\centering
	\includegraphics[scale=0.45]{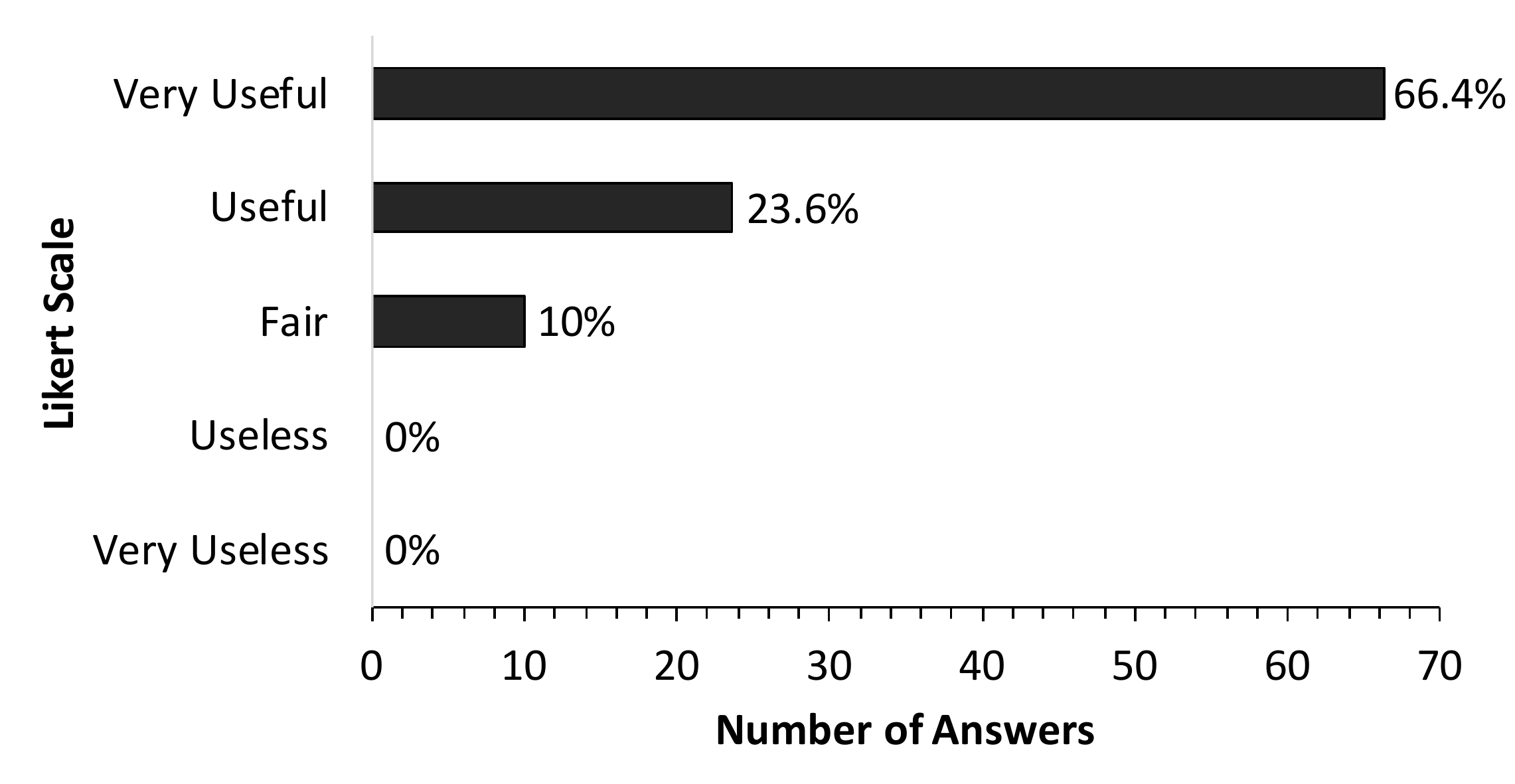}
	\caption{Usefulness of the Bot's Answers}
	\label{fig:Results on Bot's Usefulness Answers Question}
\end{figure}

Participants were asked to indicate the usefulness of the answer provided by the bot after each question they asked. The choice was on a five-point Likert's scale from very useful (meaning, the bot provided an answer they could actually act on) to very useless (meaning, the answer provided does not help answer the question at all). The participants also had other choices within the range, which were: useful (meaning, the answer was helpful but could be enhanced), fair (meaning, the answer gave some information that provided some context, but did not help the answer fully) and useless (meaning, the reply did not help with the question, but a reply was made).

Figure~\ref{fig:Results on Bot's Usefulness Answers Question} shows the usefulness results in case they were correct. Overall, \textbf{90.0\% of the participants indicated that the results returned by the bot were considered to be either useful or very useful}. Another 10.0\% indicated that the bot provided answers that were fair, meaning the answers helped, but were not particularly helpful in answering their question. It is important to emphasize that when considering usefulness, we considered all tasks that were performed correctly.

Upon closer examination of the fair results, we found a few interesting reasons that lead users to be partially dissatisfied with the answers. First, in some cases, the users found that the information returned by the bot to not be easily understandable. For example, if a user asks for all the commit logs of commits that occurred in the last year, then the returned answer will be long and terse. In such cases, the users find the answers to be difficult to sift through, and accordingly indicate that the results are not useful. Such cases showed us that perhaps we need to pay attention to the way that answers are presented to the users and how to handle information overloading. We plan to address such issues in future versions of our bot framework. Another case is related to information that the users expected to see. For example, some users indicated that they expect to have the commit hash returned to them for any commit-related questions. Initially, we omitted returning the commit hashes (and generally, identification info) since we felt such information is difficult to read by users and envisioned users of the bot to be more interested in summarized data (e.g., the number of commits that were committed today). Clearly, the bot proved to be used for more than just summarized information and in certain cases users were interested in detailed info, such as a commit hash or bug ID. All of these responses provided us with excellent ideas for how we will evolve the bot.

\conclusion{The majority (90.0\%) of the bot's users found it to be useful or very useful. Areas for improvement include figuring out how to effectively present the bot's answers to users.}

\subsection{RQ2: How quickly can users complete their tasks using the bot?}
\begin{figure}[t!]
	\centering
	\includegraphics[scale=0.54]{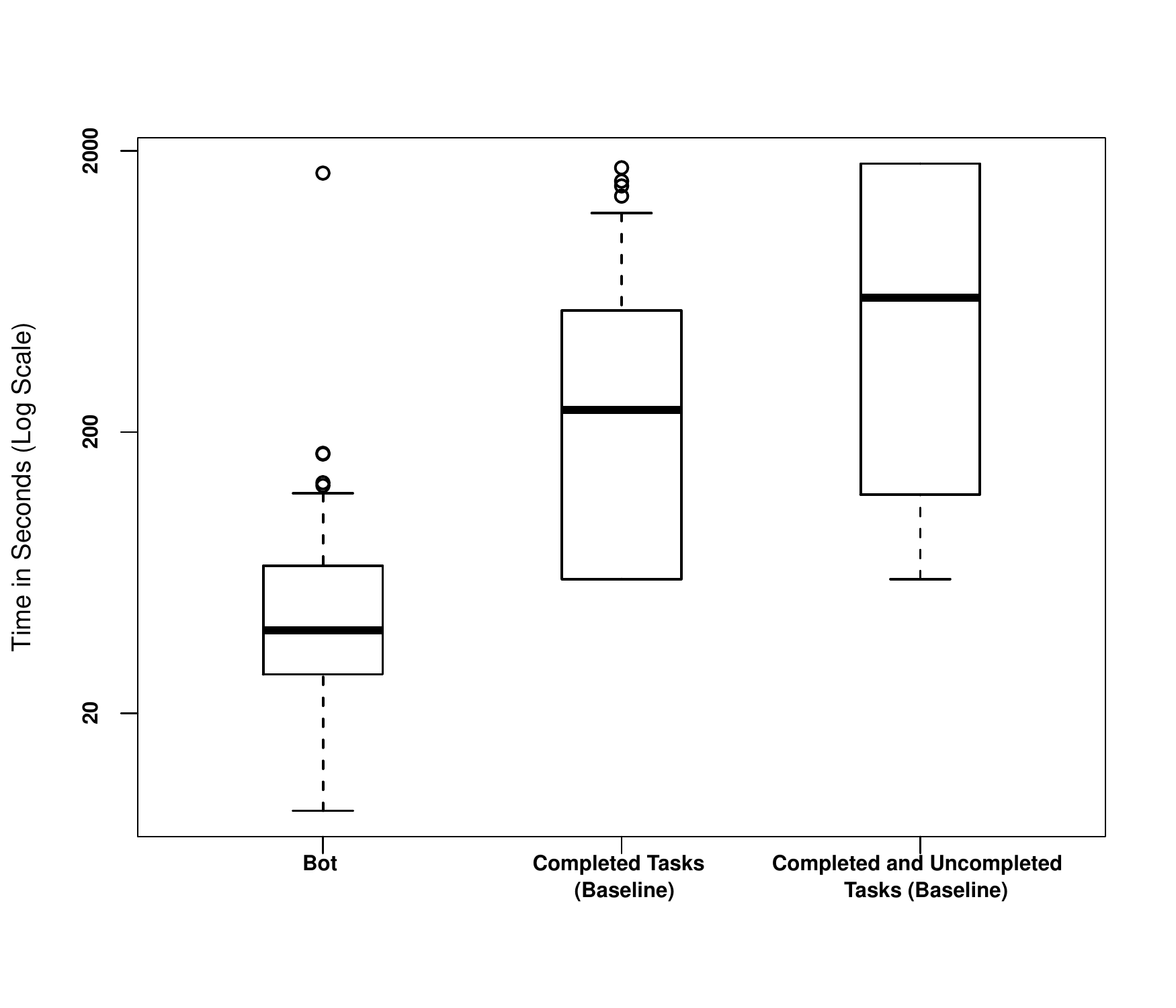}
	\caption{Time Required to Complete Tasks (bot vs. baseline)}
	\label{fig:Box Plot Speed}
\end{figure} 
Since bots are meant to answer questions in a chat-like forum, speed is of the essence. Therefore, our second RQ aims to shed light on how fast can users perform tasks related to their repositories using the bot and compare that to the time they need to complete the same tasks without the bot (i.e., the baseline). We also ask the users to indicate their perceived speed of the bot.

\noindent\textbf{Measured speed.} We measure the exact time that the participants needed to complete each of the given tasks, once using the bot and another without the bot, which we call the baseline. This gives us better insights about the actual time savings when using the bot.

Figure~\ref{fig:Box Plot Speed} shows boxplots of the distribution of the time it took for the participants to perform the tasks, with and without the bot (note that the y-axis is log-scaled to improve readability). As evident from Figure 5, using the bot (the left most box plot) significantly outperforms the baseline approach, achieving a median task completion time of 40 seconds and a maximum of 1666 seconds. On the other hand, for the baseline approach, we have two results - one that considers all tasks that users were able to complete (labeled ``Completed Tasks (Baseline)'' in Figure 5) and the other considering all tasks, i.e., completed and uncompleted\footnote{Since we gave a maximum of 30 minutes for participants to complete a task, tasks that were not answered after 30 minutes were considered to be incomplete and also to have taken 30 minutes.} (labeled ``Completed and Uncompleted Tasks (Baseline)'' in Figure 5). The median task completion time for the tasks in the baseline approach is 240 seconds with a maximum of  1,740 seconds. While, if all the completed and uncompleted tasks are considered, the time to perform a task is even higher, with a median of 600 seconds and a maximum of 1,800 seconds. To ensure that the difference between the bot and the two cases of the baseline is statistically significant, we performed a Wilcox test, and the difference in both cases (i.e., using the bot vs. completed tasks in the baseline and using the bot vs. all tasks in the baseline), and find that the difference is statistically significant (i.e., p-value$ \le 0.01$). It is obvious that using the bot to perform tasks is faster than the baseline approach. However, this research question investigates the amount of time that the bot saves compared to users manually doing the tasks, which in our case is more than 3 minutes/task, on median.

\begin{figure}[t!]
	\centering
	\includegraphics[scale=0.45]{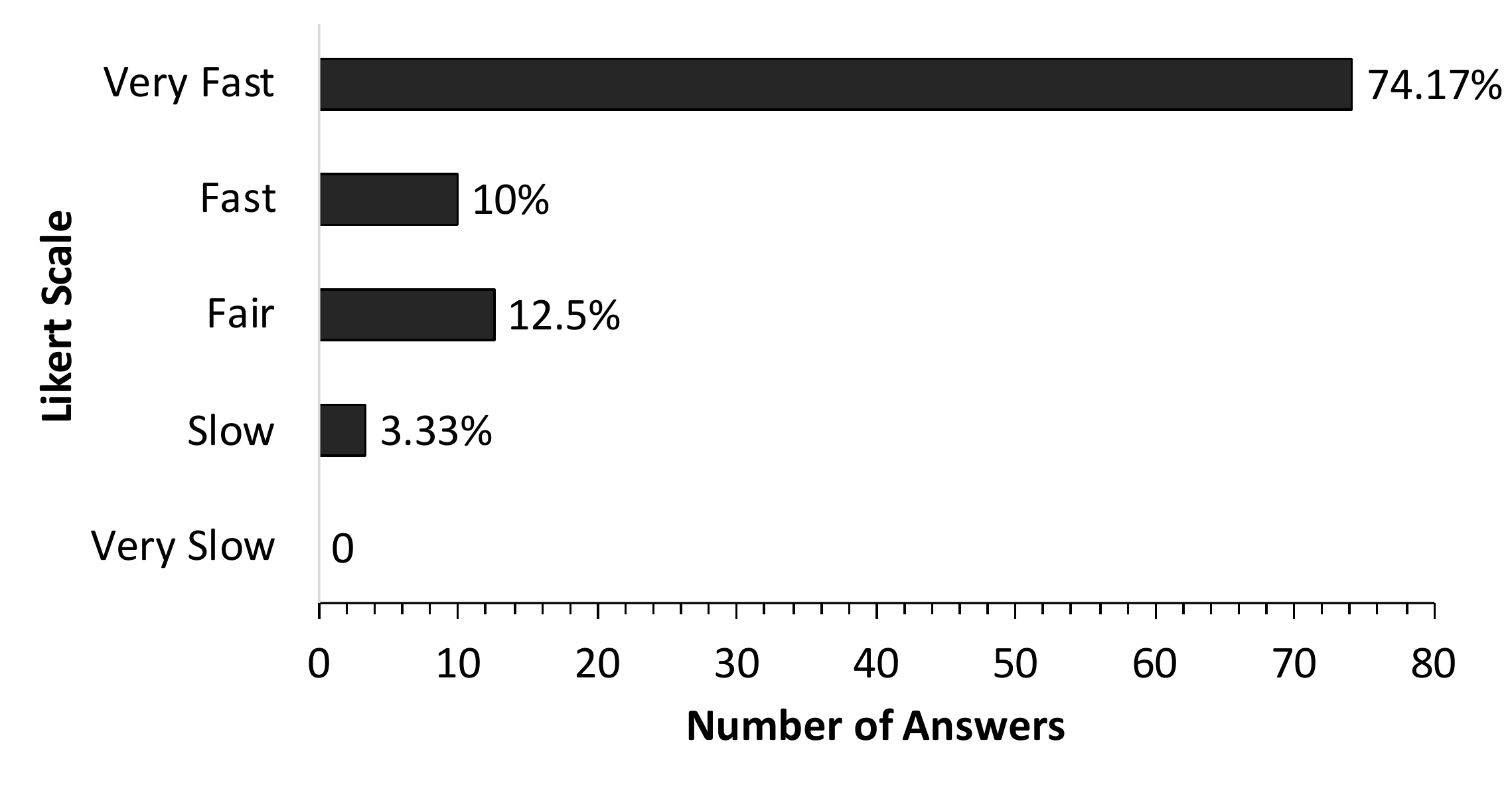}
	\caption{Speed of the Bot's Reply}
	\label{fig:Results on Bot's Replying Speed Question}
\end{figure} 

\noindent\textbf{Perceived speed.} The other side of the coin is to determine how users perceive the speed of the bot to be. To accomplish this, we asked users to indicate how fast they received the answer to their question from the bot. Once again, the choices for the users were given on a five point Likert's scale, from very fast (approx. 0 - 3 seconds) to very slow ($\geq$ 30 seconds). The participants also had other choices within the range, which were: fast (4 - 10 seconds), fair (11 - 20 seconds) and slow (21 - 30 seconds). 

Figure~\ref{fig:Results on Bot's Replying Speed Question} shows the results of the survey participants. The majority of the responses (84.17\%) indicated that the bot's responses were either, fast or very fast. The remaining 15.83\% of the replies indicated that the bot's response was either fair or slow. Clearly, our answers show that the bot provides a significant speed up to users.

\noindent\textbf{Deep-dive analysis.} To better understand why some of the tasks took longer to perform using the bot, we looked into the logged data and noted 4 cases that may have impacted the response speed of the bot. We found that in those cases, Dialogflow took more than 5 seconds to extract intents and entities from the user's question. We searched for the reasons behind Dialogflow's delay and found that the way users ask questions can make it difficult for Dialogflow's algorithms to extract the entities and intents. On the other hand, there is one case where the participant required more than 30 minutes to complete their task using the bot. We followed up with the participant afterwards to determine the reason and were told that the participants simply ``forgot'' to complete the task since they were distracted.

As for the case where users took a long time to find that answers in the baseline case, we found that the main reason for such delays is that some tasks were more difficult to answer. Hence, users needed to conduct online searches (e.g., using Stack Overflow) of ways/techniques that they can use to obtain the answer.

That said, overall, the participants were fast in completing tasks using the bot. It is important to keep in perspective how much time using the bot saves. As we learned from the feedback of our baseline experiments, in many cases, and depending on the task being performed, a developer may need to clone the repository, write a short script, and process/clean-up the extracted data to ensure that they complete the tasks correctly - and that might be a best case scenario. If the person looking for the information is not very technical (e.g., a manager), they may need to spend time to learn what commands they need to run or tools to use, etc., which may require several hours or days.

\conclusion{The participants take a median time of 40 seconds to perform a task using the bot. Moreover, the majority (84.17\%) of the bot's users perceived the bot's responses to be fast or very fast. However, the way that the user frames the question may impact the speed of the bot's reply.}

\subsection{RQ3: How accurate are the bot's answers?}
In addition to using the typical measures to evaluate bots, i.e., usefulness and speed, it is critical that the bot returns accurate results. This is of particular importance in our case, since software practitioners generally act on this information, sometimes to drive major tasks.

\noindent\textbf{Bot's performance.} We measure accuracy by checking the tasks' results performed by the users using the bot and comparing it with the actual answer to the task if it was queried manually by cloning the repositories then writing a script to find the answer or executing  git/Jira commands. For example, to get the developers who touched the "KafkaAdminClient" file, we ran the following git command: "git log --pretty=format:\%cn -- clients/src/main/java/org/apache/kafka/clients/admin/KafkaAdminClient.java". This RQ checks each component's functionality in the framework. Particularly, it checks whether the extraction of the intents and entities is done correctly from the natural language question posed by the users. Moreover, we check whether our knowledge base component queries the correct data and if the response generator produces the correct reply based on the intent and knowledge base, respectively. In total, the first two authors manually checked all the 120 completed tasks by the participants using the bot.

Our results showed that the users correctly completed 90.8\% (109 of 120) of the tasks using the bot. Manual investigation of the correct tasks showed that the bot is versatile and was able to handle different user questions related to the tasks. For example, the bot was able to handle the questions \textit{``tell me the number of commits last month''} asked by participant 1 vs. \textit{``determine the number of commits that happened in last month.''} asked by participant 2 vs. \textit{``how many commits happened in the last month''} from participant 3, which clearly have the same semantics but different syntax.

\begin{table}[t]
	\centering
	\caption{Reasons for Uncompleted Tasks by the Bot} 
	\begin{tabular}{lc}
		\hline
		\multicolumn{1}{c}{\textbf{Reason}} & \textbf{Number of Questions} \\ \hline
		Extract Intent                      & 5                      \\
		Recognize Entity                    &        5               \\
		Developer's Distraction                    &        1               \\
		Out of scope                        & 9                       \\ \hline
	\end{tabular}
\label{Number_of_unanswered_questions_per_reason}
\end{table}

Our findings indicate that the 10 tasks that the users fail to complete correctly were due to the incorrect extraction of intents or entities by our trained NLU model as shown in Table \ref{Number_of_unanswered_questions_per_reason}. For example, in one scenario the user asks ``tell me the commit info between 27th May 2018 till the end of that month?'' and our NLU model was unable to identify the entity (because it was not trained on the date format mentioned in the participant's question). Consequently, the knowledge base and the response generator components mapped the wrong entity and returned an incorrect result. Interestingly, in such cases, some of the participants had to ask the bot more than once to complete the task correctly. For example, P1 posed the following question ``how many of June 2018 bugs are fixed'' to perform task 10, the bot fails to extract the correct intent (the fixing commits that induce bugs) from the question, which lead to an incorrect reply (returned the developers that are expert in fixing bugs). Consequently, the participants rephrased the query to the bot as follows: ``show me the percentage of bugs fixes that introduced bugs in June 2018'' which allowed the bot to return the correct answer. Overall, the participants asked 1.3 question per task, on average. It is important to note that we consider the task where the participant is distracted as incomplete since s/he took more than 30 minutes. 

\noindent\textbf{Baseline performance.} As mentioned earlier, we also conducted a baseline comparison where we asked users to perform the same tasks without the bot. Figure 7 shows a break down of 1) the number of completed tasks and 2) the number of \emph{correct} tasks per completed tasks. On the positive side, we can see that the survey participants were able to provide some sort of answer for all tasks, albeit some of the tasks (e.g., T3, T6, T11 and T15) had less answers from participants. Across all tasks, the participants provided some sort of answer in 62.6\% of the cases.

However, what is most interesting is that the number of correct answers is much lower. Across all tasks, the survey participants provided the correct answer in 25.2\% of the cases. For example, for T3, T11 and T15, all of the provided answers were incorrect. On the other hand, T9's answers were all correct. 

\noindent This outcome highlights another (in addition to saving time) key advantage of using the bot framework, which is that reduction of human error. When examining the results of the baseline experiments, we noticed that in many cases participants would use a wrong command or a slightly wrong date. In other cases where they were not able to provide any answer, they simply did not have the know how or failed to find the resources to answer their question within a manageable time frame.

\conclusion{Overall, the bot achieves an accuracy of 90.8\% in answering user's questions, which is much higher than the baseline's accuracy of 25.2\%. Techniques to make the NLU training more effective can help further improve the bot's accuracy.}

\begin{figure}[t!]
	\centering
	\includegraphics[scale=0.48]{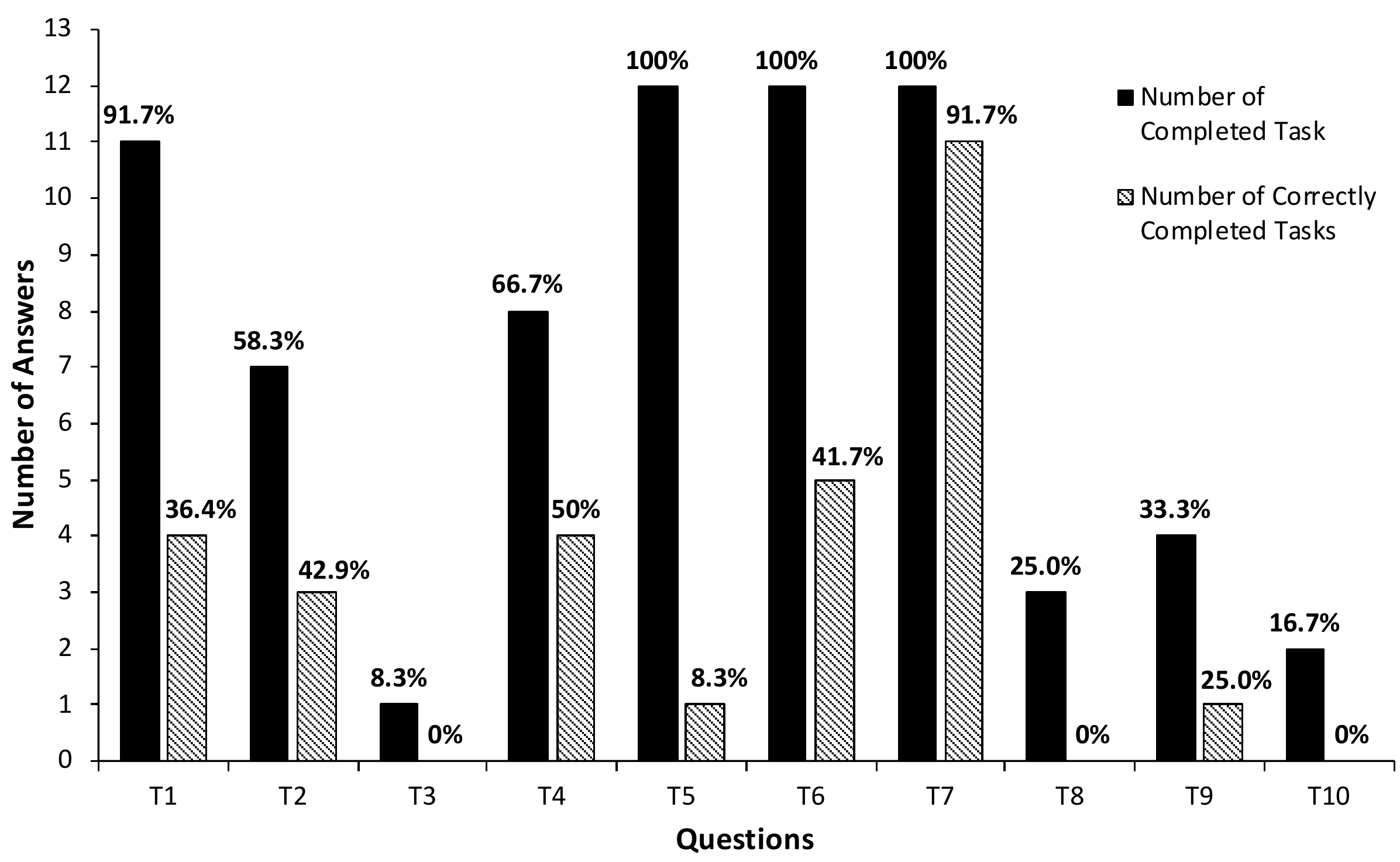}
	\caption{Number of Answers for Each Task in the Baseline}
	\label{fig:Baseline All Answers}
\end{figure}

\subsection{Follow-up Interviews}

The survey results provided us with an excellent way to quantify the usefulness, efficiency, and accuracy of the bot. However, we wanted to obtain deeper insights, particularly related to what the participants felt were the strengths and areas for improvement for the MSR-related bot framework. Therefore, we conducted semi-structured  interviews, where we sat with 5 of the 12 survey participants and asked them: 1) What they believe are the strengths of using a bot framework? 2) What they believe can be improved? and 3) any general comments or feedback they had for us? We asked for permission to record the results of the interview, to enable us to perform deeper offline analysis of the results.

In terms of strengths, most of the points mentioned during the interview surrounded the benefits and applicability of the bot on top of software repositories in industry/practical settings. We elaborate on some of the points pointed by the interviewees below:

\begin{itemize} 
\item \textbf{Bots are useful in software projects where personnel with many roles are involved.}
Although we knew from the RQs that bots will be useful, however, our interview revealed to us that bots are probably useful to more than just developers. For example, P1 and P2, two of the most experienced participants indicated that usually large projects involve personnel with varying technical backgrounds. And it is usually the less technical personnel (e.g., project managers) that can significantly benefit from repository information, but often lack the know-how to extract such information.

\item \textbf{Bots are very efficient, even when large and long-lived projects are being analyzed.}
Another major strength pointed out is the ability of the bot to provide answers from a very rich history very quickly. For example, P4 mentions the fact that some repos ``have more than 20,000 [or more] commits and if you are going through like 20,000 bugs or specific commits it will take so much time and probably you may even miss important data easily''. 

Another participant, P5 had a different perspective and saw the potential for bots to help researchers that study software projects. Particularly, he pointed out that our bot framework can be used by researchers who may want to get a few quick answers about a project or do some deep-dive analysis based on some of their findings. For example, one can ask the bot how many changes or how many bugs were reported against a file that they found to yield interesting results in their analysis.

\item \textbf{Bot's have a very low barrier-to-entry, especially since they can understand natural language.}
The participants pointed out that a major advantage is that they did not need to `learn' any specific technology or language to interact with the bot. This significantly lowers the barrier to entry for adopters of our framework. For example, P3 says ``I like that you type in natural language without thinking about building a query to do the thing [answer the question at hand]''

\end {itemize}

Other points mentioned reaffirmed our quantitative findings. For example, participants P2, P3 and P5 all mentioned the bot's speed in replying to questions and the fact that the bot helps complete the tasks at hand faster as a major benefit. We do not discuss this in detail here, since we believe our results already discussed such point and there is no point in repeating the same points again.

We also gained some valuable feedback about the potential areas of improvement for the bot framework. We elaborate on some of the points mentioned by the interviewees below:

\begin{itemize} 
\item \textbf{Support deep-dive of answers provided.}
The participants mentioned that although they appreciated the simplicity and clarity of the bot's answers, you see adding the ability of allowing the user to dive more into the results as a potential future feature. For example, participant P1 mentioned that it would be great to add hyperlinks for commit hashes or bug IDs that are returned. We believe that such a feature is indeed warranted and plan to (and believe we could easily) implement such a feature.

\item \textbf{Make the bot more resilient and modifiable.}
One clear limitation of our bot framework is that it supports a limited number of questions, even though we did that since our goal is to evaluate the viability of using a bot on top of software repositories. In any case, the participants P2 and P3 suggested that a clear improvement is to support more questions. We certainly plan to look into ways that can make our bot learn effectively from questions that are not yet supported, based on the questions users ask. Another participant, P4 suggested that we try and add a recommendation system to the bot so that typos are fixed with a ``did you mean ...'' type of messages. Also, questions that might be mistyped, e.g., how vs. who, can be provided with a suggested fix.
\end{itemize}

That said, all participants strongly showed support for the idea and mentioned they see a lot of potential for the combination of bots and software repositories. Our work is a step in the right direction, showing the value and applicability of using bots to effectively extract useful information easily from software repositories.

\section{Discussion }
\label{sec:discussion}
In this section, we present other perspectives in evaluating the software bots, and discuss the future and practical implications of our work.

\subsection{Bots Evaluation}
One of the main challenges in our work is the evaluation part since there is not much prior work that evaluates the use of bots on software repositories. One of the goals of using bots in the software engineering domain is to improve developers' productivity \cite{Storey:2016:DDP:2950290.2983989}. Therefore, we believe that any proposed bot should be evaluated against the methods that developers usually use to perform their tasks. 

As discussed in section \ref{sec:casesetup}, we evaluate the proposed framework according to its effectiveness and usefulness for developers in performing their tasks. However, there may be other ways to assess software bots. For example, usage, which can be measured by the number of times that a user uses the bot in a certain duration (e.g., a week). Other measures for speed can be related to the number of interactions needed for a user to complete a task, i.e., asking more questions to perform one task increases the time required to complete that task. 

In our case study, there are 11 cases where the participants reworded the question because the bot failed to return the correct answer because of 1) incorrect extraction of intents and entities 2) missing entities in the posed questions 3) typos in the users questions (e.g., What commits happened between 27/5/208 - 31/5/2018) 4) the bot encountered connection issues to the internet. For example, one of the participants asked the bot ``What  the details of the commits between May 27th 2018 and May 31st 2018'', the bot failed to extract the entity (May 27th 2018 and May 31st 2018) because it was not trained on date format that is specified in the posed question. Then, the user rephrased the question to the bot as follows: ``What are the details of the commit between 27/5/2018 - 31/5/2018'', which allowed the bot to extract the entity correctly. 

That said, measuring the total time needed to perform a certain task may cover the number of interactions metric. One can also argue that the number of interactions is considered as a measure for users satisfaction rather than a speed. Because a large number of interactions to perform a task impacts the satisfactions of the bot users negatively \cite{userSatisfaction_link,NegBotImpact_Report,Lebeuf_2018IST}.

In general, the evaluation of software bots varies based on their characteristics and the tasks they can perform. For example, if a bot is proactive (e.g., reminds the user to execute a certain task), then the number of interactions may be invalid as a speed metric since the conversation always comes from the bot. Storey and Zagalsky \cite{Storey:2016:DDP:2950290.2983989} suggested to measure the bot's efficiency and effectiveness in completing developers' tasks. However, we believe that there are different dimensions that bots' developers and researchers can use to assess the proposed bots regardless of their types such as accuracy and user satisfaction. For example, to measure user satisfaction, the bot can monitor and track the sentiment of a user through the conversation and take different actions based on the current user sentiment. On the other hand, bots can be evaluated based on their intelligence, such as how easily they can adapt to new contexts, their ability to explain the reasoning behind their behavior, and the degree of smoothness of the conversation flow \cite{Lebeuf_2018IST}. We believe that bots evaluation is still an active area and we encourage researchers to investigate the various dimensions and measures that the bots community can use to assessdifferent bots.

\subsection{Study Implications}

Our framework has a number of implications on both, software engineering research and practice.\newline
\\
\textbf{Implications for Future Research:} Our study mainly focused on proposing and evaluating a framework to answer some of the most common developers' questions using software repositories. Based on our results, we find a number of implications for future research. First, there is a need for the MSRBot to support more complex queries, as indicated by the participants' comments. For example, one of the participants stated that the bot is going to help a lot in answering more complex queries using the repositories data. Hence, our study motivates the need to examine more complex questions using data from different types of software repositories. Also, our study shows that there is a need to develop approaches that answers questions dynamically (i.e., removing the need to have predefined questions) using  repositories to overcome the limited number of questions supported by the bot.\\
\noindent Second, the retrieved data in the bot's answers and the way it is displayed to the user can be improved. For example, one of the participants recommends providing a suggestion to the user in case the bot was unable to identify the user's intent from the posed question. Another participant indicated that the bot's reply should be short and suggested the use of pagination to enable the user to navigate more easily through the bot's reply. Using pagination when displaying the bot's reply (e.g., commits that happened in the last 3 months) needs careful consideration since questions such as how many records to display on each page, and how can users know in which page a specific record is, need to be considered. Furthermore, it is unconventional to implement the pagination in a chat interface, which might affect user satisfaction. Although there are many studies on software bots, we are not aware of any studies that discuss the best way to represent the bot's reply and kind of information that bots' users expect to see. Our findings motivate the need for such studies.\newline

\noindent\textbf{Practical Implications:} A direct implication of our findings is that using the bot simplifies the extraction of useful information from software repositories in different ways. First, it helps project stakeholders that do not have technical skills to easily extract the information from different repositories using the natural language. And, although some developers have the skills to perform such tasks (i.e., mining and analyzing data from software repositories), our framework reduces the time to complete those tasks compared to the baseline (i.e., any tool other than MSRBot). Overall, we believe that our framework supports practitioners at different levels in performing their tasks by lowering the barrier to entry for extracting useful data from software repositories. For example, it helps project managers to track the project progress (based on the closed tickets) and assists developers in their daily tasks.

\section{Threats to validity}
\label{sec:threats}
In this section, we discuss the threats to construct, internal and external validity of our study.

\subsection{Construct Validity:} 
The 12 participants used to evaluate the bot framework may have reported incorrect results, which would impact our findings. However, we are quite confident in the results returned since 1) most of these students have professional software development experience and 2) in most cases, there was a clearly popular answer (i.e., very few outliers). We also interviewed a subset of the participants, and based on our discussions, all participants seemed very competent in evaluating the output of the bot.

We selected different questions from the literature to evaluate the proposed framework which might bias our evaluation by adding the questions that the bot can better answer. However, we mitigate this threat to validity in different ways. First, we selected the list of questions from different studies arbitrarily. Second, in the given tasks, the participants are free to ask any type of questions to the bot (that is related to the task). Lastly, the participants are unaware of the list of questions that the bot can answer or is trained on.

\subsection{Internal Validity:}
We used Google's Dialogflow to extract the intents and entities from the posed questions. Hence, our results might be impacted by Dialogflow's ability to translate the user's questions. That said, RQ3, which examines the accuracy of the bot showed high accuracy, which makes us confident in the use of Dialogflow. However, using another framework, might lead to different results. In the future, we plan to examine the impact of such frameworks on our bot. Also, Dialogflow's NLU model is trained on examples that were provided and manually labeled (i.e., the intents and entities) by us. The quality and quantity of training data may impact the effectiveness of the NLU. That said, in our evaluation, Dialogflow was able to handle the majority of questions from our users. In the future, we plan to investigate ways that our bot can learn from user's input and automate the intent and entity extraction/training phase.

Another threat to internal validity is that we used the questions identified in the literature to formulate the tasks in our case study. In some cases, the statement provided in the task might bias the participants on how to pose questions to the bot. The reason for providing the statements to the participants is that we want to keep our study manageable and to evaluate the developed bot using the supported types of questions. However, we mitigate this threat by providing the question as a statement. Also, we did not reveal the list of the questions that the bot was trained on to the participants. Albeit anecdotal, we believe that our results show a limited effect of the wording of the tasks on the participants because their questions were syntactically different than the provided statements, although they have similar semantics) (e.g., ``how many commits in the last month'' and ``what are the details of the commits between  27/5/2018 - 31/5/2018?''.

\subsection{External Validity:}
Our study was conducted using Hibernate-ORM and Kafka projects and supported 15 common questions. This means that the study might yield different results when selecting other projects or supporting a different set of questions related to software repositories. These are threats to external validity as they may limit the generalisability of our results. However, we plan to expand our study to support more systems and questions in the future. Also, it is important to note that the point of this paper is to design and evaluate the feasibility of using bots on top of software repositories to automate the answering of commonly asked questions.

Also, we used students to evaluate our framework. Therefore, using different participants (i.e., developers) might affect the generalizability of our results. However, we argue that the main purpose of the case study is to evaluate our approach rather than studying the participants characteristics. Moreover, most of the participants have more than 3 years of industrial experience which reduces the threats to external validity. Furthermore, we plan to conduct a large-scale study in the future by having more developers from the industry.

\section{Conclusion \& Future Work}
\label{sec:conclusions}

Software repositories contain numerous amounts of useful information to enhance software projects. However, not all project stakeholders are able to extract such data from the repositories since it requires technical expertise and time. In this paper, we design and evaluate the feasibility of using bots to support software practitioners in asking questions about their software repositories. Our findings show that the bot provides answers that are considered very useful or useful (as indicated by 90\% of the participants), efficient (participants take on median 40 seconds to complete the tasks), and accurately answers questions posed by its users (as indicated by 90.8\% of completed tasks). Also, our study highlighted some of the potential pitfalls when applying bots on software repositories. For example, our study finds that more attention is needed regarding the content and length of the information that the bot replies with, how to handle complex user questions, and how to handle user errors such as typos. Overall, our work showed that bots have the potential of playing a critical role by lowering the barrier-to-entry for software stakeholders to extract useful information from their repositories. Also, the bots are efficient when used on large projects with a long history, which saves the developers' time needed to extract the data from the repositories. Finally, we believe that our work encourages other researchers to explore different usages of bots on different types of software repositories, e.g., code review repositories.
	
In addition to addressing the issues found in the user study (e.g., handling user typos), the results in this paper outline some directions for future work. First, since the entity recognizer and intent extractor components accuracy depend on the training dataset, we will examine different techniques to generate a dataset for those components (e.g., using Stack Overflow topics) to increase their accuracy. We want to compare the performance of different Natural Language Understanding Platforms (e.g., Dialogflow and Rasa) using software engineering datasets to help the bot's community identify the platform that best fits their context. Also, we are planning to support a wider range of repositories such as (e.g., Gerrit Code Review). Moreover, in the current implementation of our framework, we did not provide the users with the ability to configure the bot. However, allowing users to configure the bot may improve the flexibility and adaptability of our framework. In the future, we plan to allow the MSRBot to be configured through users' feedback by asking users for their preferences and suggesting possible configurations. Finally, we plan to evaluate our bot framework using industrial settings to gain more insights into the roles and scenarios in which the bot can be used.

\bibliographystyle{abbrv}

\bibliography{bots_bib} 

\end{document}